\documentclass[aps,prl,twocolumn,showpacs]{revtex4}  
\usepackage{epsfig}
\begin{document}
 \title{Edge spin accumulation in a ballistic regime}
 \author{Alexander Khaetskii$^1$, and Eugene Sukhorukov$^2$}
\affiliation{$^1$ Institute of Microelectronics Technology, Russian Acedemy of Sciences, 142432 Chernogolovka,
Moscow District, Russia}
\affiliation{$^2$ Department of Theoretical Physics, University of Geneva,
24 quai Ernest Ansermet, CH-1211, Switzerland}

\date{\today}


\begin{abstract}
We consider a mesoscopic {\it ballistic} structure with Rashba spin-orbit splitting of the electron spectrum. The ballistic  region is attached to the leads with a voltage applied between them. We calculate the edge spin density which appears in the presence of a charge current through the structure due to the difference in populations of electrons coming from different leads. 
{\r Combined effect of the boundary scattering and spin precession leads to} oscillations of the edge polarization with the envelope function decaying as a power law of the distance from the boundary. The problem is solved with the use of scattering states. The simplicity of the method allows to gain an insight into the underlaying physics. We clarify the role of the unitarity of scattering for the problem of edge spin accumulation. In case of a straight boundary it leads to exact cancellation of all long-wave oscillations of the spin density. As a result, only the Friedel-like spin density oscillations {\r with the momentum} $2k_F$ survive. However, this appears to be rather exceptional case. In general, the smooth spin oscillations with the spin precession length recover, as it happens, e.g., for the wiggly boundary. We demonstrate also, that there is no relation between the spin current in the bulk, which is zero in the considered case, and the edge spin accumulation.  
\end{abstract}

\maketitle

   Currently there is a great interest, both experimental and theoretical, to spin currents and spin accumulation in various mesoscopic semiconductor structures \cite{Rashba,We}. Both phenomena  are due to spin-orbit coupling and are  of great importance for future spin electronics. The edge spin density accumulation related to the Mott asymmetry in scattering off impurities was recently measured \cite{Kato}.  
  Moreover, the edge spin density in the 2D hole system which is due to the so-called intrinsic mechanism of the spin-orbit interaction was also observed \cite{Wunderlich}.  
  One of the question of the interest is the relation between the spin current in the bulk and the spin density accumulated near the sample boundary. 
     It is well known that in the diffusive regime (when a spin diffusion length is much larger than a mean free path), such a relation indeed exists \cite{We}, and the spin density appearing near the boundary is entirely determined by the spin flux coming from the bulk.  For example, in the diffusive regime and in the case of Rashba Hamiltonian, when the spin current in the bulk is zero, the spin density component perpendicular to the plane is zero everywhere down to the sample boundary \cite{BC}. 
    \par
     However, in an opposite case, when the spin precession length is much shorter than the mean free path, the situation is much less clear.  
     An example of such a situation is a mesoscopic ballistic (or quasi-ballistic) structure with spin-orbit-related splitting of the electron spectrum characterized by the energy  $\Delta_R$, when one has the limit $\Delta_R \tau_p \gg 1$, where $\tau_p$ is the mean free time. Besides, in the presence of spin-orbit interaction, the boundary scattering itself is the source of generation of the spin density. It is quite clear that the characteristic length near boundary where the spin density appears is the spin precession length, $L_s=\hbar v_F/\Delta_R$, with $v_F$ being the Fermi velocity. Thus in general there are two sources of the edge spin accumulation, the spin generation by the boundary scattering itself and the one due to the spin flux coming from the bulk (as it should be, at least, in the case of 2D holes with a qubic s-o Hamiltonian). It is a separate problem to find the relative contributions of these two mechanisms to the edge spin accumulation. Now we study only the first mechanism when the spin density is generated upon the boundary scattering. 
\par
\begin{figure}
\begin{center}
\epsfig{file=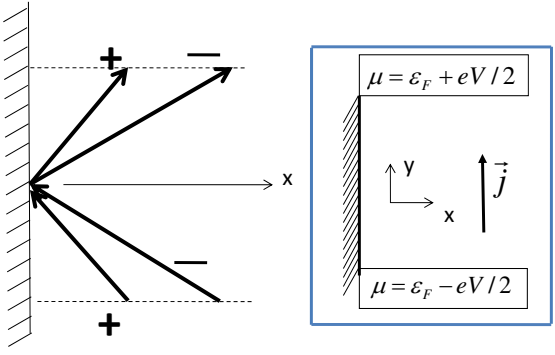,width=0.45\textwidth}
\end{center}
\caption{Left: Schematics of the boundary specular scattering in the presence of spin-orbit coupling. Plus and minus modes are shown for the same energy and the same wave vectors along the boundary. Right: Geometry of the system. Voltage $V$ applied to the ideal leads causes a charge current through the ballistic region.}
\label{fig:Spin1}
\end{figure}

For that we consider a mesoscopic  structure described by Rashba Hamiltonian  in 
the {\it ballistic} limit when a mean free path  is much greater than the sample sizes. The ballistic  region is attached to the leads, and a voltage $V$ applied between the leads  causes a charge current through the structure, as shown in Fig.\ \ref{fig:Spin1}.  
We stress that the electric field is absent inside an ideal ballistic conductor, therefore the edge spin polarization appears not as a result of acceleration of  electrons by an electric field and associated spin precession. Instead, the edge spin accumulation is due to the difference in populations of left-moving and right-moving electrons.  Combined effect of the boundary scattering and spin precession leads to oscillations of the edge polarization. Thus, there is no relation between the spin current in the bulk, which is zero in the case considered, and the edge spin accumulation.  
 \par
  We solve the problem by using scattering theory, with scattering states coming from different leads of the structure and, therefore, having different occupations. The simplicity of the method allows to gain an insight into the underlaying physics. The problem of the spin density accumulation in the ballistic case and for the straight boundary was  erlier considered in Ref.\ [\onlinecite{Zyuzin}] with the help of the Green's functions method (see also Ref.\ [\onlinecite{Reynoso}]).
Surprisingly, in this case all the long-wave oscillations of the spin density cancel,  and the final result contains only the Friedel-like oscillations with  the momentum $2k_F$.   Unfortunately, the method used in Ref.\ [\onlinecite{Zyuzin}] does not allow a simple interpretation of the puzzling result.  
\par
Here we show that it is the unitarity of 
scattering that leads to exact cancellation of the long-wave oscillations  of the spin density with the $L_s$ period  in case of a straight boundary. It should be also mentioned that the observed behaviour is closely related to the effective one-dimensional character of the scattering problem arising from the translational invariance along the boundary.  This result may be interpreted as spin-orbit splitting of the Friedel oscillations in the charge density. In other words,  two charge oscillations corresponding to spin-up and spin-down orientations, which coincided without s-o coupling, get shifted with respect to each other in the presence of the s-o interaction.  Thus, short-wave length oscillations acquire in addition smooth envelope function with the characteristic length $L_s$ \cite{Zyuzin}.  
Therefore, strictly speaking,  this phenomenon is different from a spin-orbit-related  accumulation of the spin density upon boundary scattering.

\par
However, the case of a straight boundary appears to be a rather exceptional case, while in general the smooth spin oscillations with the spin precession length  $L_s$ recover, as it happens, for example, for the wiggly boundary, or for scattering off a circular impurity in a 2D electron system \cite{2Dcase}. This is a consequence of the fact that in higher dimensions the conditions of the unitarity of scattering take a different form, as explained below.  In all these situations the spin density decays as a power law of the distance from the scatterer.

    \par 
    {\it The straight boundary.} 
       The s-o Hamiltonian {\r in the bulk of a ballistic 2D electron system takes the following form}
\begin{equation}
\hat {\cal H}({\bf p})=\frac{p^2}{2m}+ \frac{\alpha}{2}\vec{n}[\vec{\sigma}\times {\bf p}],
\,\,\,  \epsilon_M(p)= \frac{p^2}{2m}+ \frac{M}{2}\alpha p,
\label{Rashba}
\end{equation}
where $\vec{n}$ is the normal to the plane, ${\bf p}$ is the 2D momentum, and $ M=\pm $ are the helicity values.  
The solutions of this Hamiltonian have the form $\exp (i{\bf p}{\bf r}/\hbar)\chi_M({\bf p})$, where ${\bf r}=x,y$ and 
the explicite form of the spinors is
$$\chi_{\pm}(\varphi)= \frac{1}{\sqrt{2}}\left( 
\begin{array}{ll}
1 & \\
\mp i e^{i\varphi} &
\end{array}
\right)$$
with $\varphi$ being the angle between the momentum ${\bf p}$ and the positive direction of the $x$-axis. 
Now we consider the semi-infinite system and choose the $x$-axis to be directed perpendicular to the boundary ($x=0$) of the 2D system, see Fig.\ \ref{fig:Spin1}. The wave functions which obey zero boundary conditions 
at $x=0$ are obviously the {\it scattering states} which are the superposition of the bulk solutions and constitute the complete set of the orthonormal functions.  
Two scattering states corresponding to incident plus and minus modes with given wave vector along the boundary and the same energy are
\widetext
\begin{eqnarray}
\hat{\Psi}_{+}^{(0)}(x,y)=e^{ik_yy}[\chi_{+}(\pi -\varphi)e^{-ikx}+ F_{+}^{+}\chi_{+}(\varphi)e^{ikx}+               F_{+}^{-}\chi_{-}(\varphi_1)e^{ik_1x}]; \,\,\,  \hat{\Psi}_{+}^{(0)}(0,y)=0, 
\label{plusmode} \\
\hat{\Psi}_{-}^{(0)}(x,y)=e^{ik_yy}[\chi_{-}(\pi -\varphi_1)e^{-ik_1x}+ F_{-}^{+}\chi_{+}(\varphi)e^{ikx}+               F_{-}^{-}\chi_{-}(\varphi_1)e^{ik_1x}]; \,\,\,   \hat{\Psi}_{-}^{(0)}(0,y)=0.
\label{minusmode}
\end{eqnarray}
\endwidetext
 The wave vectors are defined in the following way 
 \begin{equation}
 k^2=k_+^2-k_y^2, \,\,\, k_1^2=k_-^2-k_y^2, \,\,\, \hbar k_{\pm}=m(v_F \mp \frac{\alpha}{2}),
 \label{k_vectors}
 \end{equation}
 where $\hbar k_{\pm}$ are the momenta at the Fermi energy in the plus and minus modes.  The angles $\varphi$, $\varphi_1$  are defined via $\sin \varphi=k_y/k_+$ and $\sin \varphi_1= k_y/k_-$, as indicated in Fig.\ \ref{fig:Spin1}.  
 From Eqs.(\ref{plusmode}) and (\ref{minusmode}) one finds the scattering amplitudes $F_{+}^{+}$ and $F_{+}^{-}$:
\begin{equation}
F_{+}^{+}=-\frac{(e^{i\varphi_1}-e^{-i\varphi})}{(e^{i\varphi_1}+e^{i\varphi})}; \,\,\, F_{+}^{-}= -\frac{2\cos\varphi}{(e^{i\varphi_1}+e^{i\varphi})}.
\label{ScattAmpl}
\end{equation}
 One can check that the amplitudes $F_{-}^{-}$ and  $F_{-}^{+}$ for the incident minus mode with the same $k_y$ and the same energy  are obtained from $F_{+}^{+}$ and  $F_{+}^{-}$ by the replacement $\varphi \leftrightarrow  \varphi_1$. The components of the unitary scattering matrix $\hat{S}$ have the following form:
  \begin{equation}
  S_+^+=F_{+}^{+},\,\, S_-^-=F_{-}^{-}, \,\, S_+^-= S_-^+=F_{+}^{-}\sqrt{\frac{v_{x,-}}{v_{x,+}}},
  \label{SMatrix}
  \end{equation}
 where $v_{x,i}=\partial \epsilon_i(p)/\partial p_x$ are the group velocities.  Note that for the Rashba model one has 
 $v_{x,-}/v_{x,+}= \cos\varphi_1/\cos\varphi$. 
 \par 
   The wave functions Eqs.\ (\ref{plusmode}) and (\ref{minusmode}) may now be used to calculate the average $z$-component of the spin, $S_z$, as a function of coordinates:
  \begin{equation}
  <S_z(x)>=\sum_{i=\pm}\int \frac{dk_y}{(2\pi)^2}\frac{d\epsilon}{v_{x,i}}f_F(\epsilon,k_y) 
  <\hat{\Psi}_{i}^{(0)}(x)|\hat S_z |\hat{\Psi}_{i}^{(0)}(x)> 
  \label{S_z}
\end{equation}
 where $f_F(\epsilon,k_y)$ is the Fermi distribution function of the lead with the chemical potential shifted by $\pm eV/2$. Depending on the sign of $k_y$, this function takes either of two values, 
  $f_F(\epsilon -\mu -eV/2)$ or $f_F(\epsilon -\mu +eV/2)$. 
  Substituting scattering states (\ref{plusmode}) and (\ref{minusmode}) to Eq.\ (\ref{S_z}), we find that 
  one may distinguish various contributions to $<S_z(x)>$ with different oscillation periods, which originate from an interference of different terms in  Eqs. (\ref{plusmode}) and (\ref{minusmode}). 
  The part of $<S_z(x)>$  which involves the interference of the outgoing waves (two last terms in Eqs.\ (\ref{plusmode}),(\ref{minusmode})), reads:
  \begin{eqnarray}
  \int dk_y d\epsilon f_F(\epsilon,k_y) \frac{1}{\sqrt{v_{x,-}v_{x,+}}} \nonumber \\ \times
  \Bigg [A<\chi_{-}(\varphi_1)|\hat{S}_z|\chi_{+}(\varphi)>e^{i(k-k_1)x} +{\rm c.c.}\Bigg ],
    \label{Unitary}
\end{eqnarray}
where 
$$
 A=S_{+}^{+}\cdot (S_{+}^{-})^{\star} + S_{-}^{+} \cdot (S_{-}^{-})^{\star}.
 $$
Here we used the fact that the distribution function $f_F(\epsilon,k_y)$ has the same value at given energy 
  for  both plus and minus modes. 
    Note, that the period of oscillations of the exponential factor $e^{i(k-k_1)x}$ in Eq.\ (\ref{Unitary}) is of the order of the spin precession length. 
   However, the term (\ref{Unitary}) vanishes, because the expression $A$ is exactly equal to zero. Indeed, this expression is nothing but a non-diagonal component of the identity matrix $\hat{S}\hat{S}^{\dagger}$. 
   Thus, we obtain an interesting result that the only reason for the absence of the long-wave oscillations in  $<S_z(x)>$ is the unitarity of scattering.
          
  Finally, by taking  into account in Eq.\ (\ref{S_z}) the terms responsible for the interference between incoming and the outgoing waves (for example, between first and second terms in Eq.\ \ref{plusmode}), and adding the contribution from the evanescent modes \cite{evanes}, we reproduce Eq.\ (16) of the Ref.\ \cite{Zyuzin}.  There it has been shown that so derived spin density may be approximated by the expression $S_z(x)\propto eV \cos(2k_Fx)\sin^2(x/L_s)/x$.  Therefore, the total spin per unit length along the boundary scales as $\int_0^{\infty}dx S_z(x) \propto \alpha^2$. Note, that the main contribution to this integral comes from small distances from the boundary, $x\simeq\lambda_F$, which makes it to depend strongly on the boundary quality. In what follows we show that, indeed, breaking the translation invariance along the boundary induces long wave length oscillations of the spin density and leads to the total spin that is not small in the parameter $\alpha$.
   \par
                  
\begin{figure}
\begin{center}
\epsfig{file=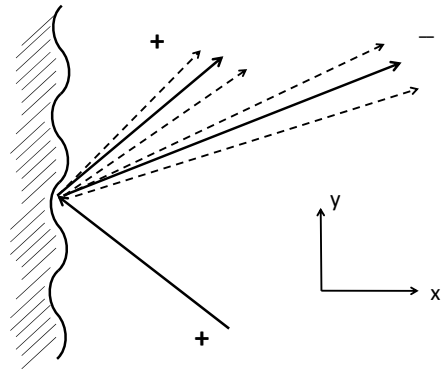,width=0.4\textwidth}
\end{center}
\caption{Schematics of the scattering of the plus incident mode by the wiggly boundary, $x=W\sin(2\pi y/\lambda)$. Besides the main scattering channels (solid lines), one has additional scattering waves with the wave vectors along the boundary shifted by $\pm 2\pi/\lambda$ (dashed lines).}
\label{fig:Spin2}
\end{figure}

      {\it Scattering by the wiggly boundary.} We consider here scattering off the wiggly boundary, shown in Fig.\ \ref{fig:Spin2}. Because in this case the translational invariance is broken,  the condition of the unitarity of scattering takes a different form as compared to the case of the straight boundary, and, as a result, the cancellation of the smooth spin density oscillations does not take place. In order to demonstrate this effect we consider the mathematically simple case of the abrupt impenetrable boundary described by the equation: $x=\zeta(y)\equiv W\sin (2\pi y/\lambda)$. To the lowest order in $W$ the boundary condition reads:
      \begin{equation}
     \hat{\Psi}(0,y)+ \zeta(y)\frac{d\hat{\Psi}(0,y)}{dx}=0
     \label{bouncond}
     \end{equation}
   We are looking for the solution in the perturbative form $ \hat{\Psi}_{\pm}(x,y)= \hat{\Psi}_{\pm}^{(0)}(x,y)+ \hat{\Psi}_{\pm}^{(1)}(x,y)$, where the first order correction is proportional to $W$, and the zeroth order functions are given by Eqs.(\ref{plusmode}),(\ref{minusmode}). 
    Using that $\hat{\Psi}_{\pm}^{(0)}(0,y)=0$, we obtain
    \begin{equation}
     \hat{\Psi}_{\pm}^{(1)}(0,y)+ \zeta(y)\frac{d\hat{\Psi}_{\pm}^{(0)}(0,y)}{dx}=0
     \label{bouncond1}
     \end{equation}
The following wave functions satisfy the above equation: 
    \widetext
    $$\hat{\Psi}_{-}^{(1)}(x,y)= ae^{i2\pi y/\lambda} e^{ik_1^>x} \chi_-(\varphi_1^>) +
    be^{i2\pi y/\lambda} e^{ik^>x} \chi_+(\varphi^>)+ ce^{-i2\pi y/\lambda} e^{ik_1^<x} \chi_-(\varphi_1^<)
    + de^{-i2\pi y/\lambda} e^{ik^<x} \chi_+(\varphi^<),
     $$
    $$\hat{\Psi}_{+}^{(1)}(x,y)= \tilde{a}e^{i2\pi y/\lambda} e^{ik^>x} \chi_+(\varphi^>) +
    \tilde{b}e^{i2\pi y/\lambda} e^{ik_1^>x} \chi_-(\varphi_1^>)+ \tilde{c}e^{-i2\pi y/\lambda} e^{ik^<x} \chi_+(\varphi^<)
    + \tilde{d}e^{-i2\pi y/\lambda} e^{ik_1^<x} \chi_-(\varphi_1^<),
     $$
     \endwidetext
    with the coefficients $a,b$, etc., being easily found from Eqs.\ (\ref{plusmode}), (\ref{minusmode}), and (\ref{bouncond1}). 
     The definitions of the k-vectors are
    $$
    k^{\frac{>}{<}}=\sqrt{k_+^2-(k_y \pm \frac{2\pi}{\lambda})^2}, \,\,\,  k_1^{\frac{>}{<}}=\sqrt{k_-^2-(k_y \pm \frac{2\pi}{\lambda})^2},
    $$
      and angles are defined as $\varphi_1^{>}=\varphi_1+\delta \varphi_1^{>}, \varphi_1^{<}=\varphi_1+\delta \varphi_1^{<}$, and so on.
     To the first order in $1/\lambda k_F$, 
    \begin{equation}
    \delta \varphi_1^{\frac{>}{<}}=\pm 2\pi/\lambda k_1, \,\, 
     \delta \varphi^{\frac{>}{<}}=\pm 2\pi/\lambda k. 
     \label{delta}
    \end{equation}
    These equations are valid under the condition $(2\pi/\lambda k_F)\ll \cos^2 \varphi$.   
    \par
     {\it Calculation of $<S_z>$.}
    To simplify somewhat long expressions, from now on we set $1/\lambda =0$ in the pre-exponential factors and keep only the terms of the lowest order in $\alpha$. 
    In particular, in view of Eq.\ref{delta} we disregard the difference between $\varphi^>$ and $\varphi$, as well as between $\varphi_1^>$ 
    and $\varphi_1$, etc., in the expressions for spinors. In other words, we assume  $\lambda_F \ll L_s \ll \lambda$.   Then, the first order correction for $<S_z>$ in the minus mode reads
   \widetext
   \begin{eqnarray}
   <\hat{\Psi}_- |\hat S_z |\hat{\Psi}_- > = 
    \left <(ae^{i2\pi y/\lambda} e^{ik_1^>x} +
  ce^{-i2\pi y/\lambda} e^{ik_1^<x}) \chi_-(\varphi_1)|\hat S_z |F_-^+ \chi_+(\varphi)e^{ikx}+ \chi_-(\pi -\varphi_1)e^{-ik_1x}\right > +  \nonumber \\
  + \left <(be^{i2\pi y/\lambda} e^{ik^>x}
    + de^{-i2\pi y/\lambda} e^{ik^<x}) \chi_+(\varphi)|\hat S_z |F_-^- \chi_-(\varphi_1)e^{ik_1x} + \chi_-(\pi -\varphi_1)e^{-ik_1x}\right > + c.c. 
  \end{eqnarray}
  \endwidetext
   One can easily  see that the contribution to $<S_z>$ from the plus mode differs from the contribution from
   the minus mode by replacing everywhere $k \leftrightarrow k_1$ and  $\varphi \leftrightarrow  \varphi_1$. 
Thus we arrive at the following expression for the quantity in question:
   \widetext
   \begin{eqnarray}
  <S_z(x,y)> = \frac{eV\cdot W}{(2\pi)^2 m v_F^2}\int_0^{k_+}dk_y k_y  \Bigg\{ (k+k_1)\sin(k_1x) [\cos(k^>x+\frac{2\pi y}{\lambda})-  \cos(k^<x-\frac{2\pi y}{\lambda})]
  \nonumber \\ -k_1\left [ \sin[(k-k_1^>)x -\frac{2\pi y}{\lambda}]- \sin[(k-k_1^<)x +\frac{2\pi y}{\lambda}] 
   + \sin[(k_1+k_1^>)x +\frac{2\pi y}{\lambda}]-\sin[(k_1+k_1^<)x -\frac{2\pi y}{\lambda}]\right] +(k \leftrightarrow k_1) \Bigg\} 
    \label{propagating}
   \end{eqnarray}
   \endwidetext
   Here the operation $(k \leftrightarrow k_1)$ should be understood as follows: $k \rightarrow k_1, \,\, k_1^> \rightarrow k^>$, etc.  Note that this equation  takes into account only the propagating modes. It means that $k_y<k_+$, and for both plus and minus modes the wave vectors in x-direction are real. If, on the other hand, $k_+<k_y<k_-$,  then for the given energy  in the plus mode the wave vectors in x-direction are purely imaginary. Those modes are called evanescent.  We will see that in contrast to the case of straight boundary there is no exact cancellation of the contribution of the propagating and evanescent modes into the edge spin density with $\xi$ oscillations (see below).   
   We present here the expression for $<\Psi_- |\hat S_z |\Psi_->$ which is due to the evanescent modes in the case when all quantities $\kappa=\sqrt{k_y^2-k_+^2}$, $\kappa^{\frac{>}{<}}=\sqrt{(k_y \pm 2\pi/\lambda)^2-k_+^2}$ are real
   \widetext
   \begin{eqnarray}
   <\Psi_- |\hat{S}_z |\Psi_->= \frac{Wk_y}{k_F}\cos \varphi_1 \sin(k_1x)[k_1\cos (\frac{2\pi y}{\lambda})(e^{-\kappa^> x}-e^{-\kappa^< x})- 
   \kappa \sin (\frac{2\pi y}{\lambda})(e^{-\kappa^> x}+e^{-\kappa^< x})] \nonumber \\
  + \frac{Wk_yk_1}{k_F}\cos \varphi_1 
   \{ e^{-\kappa x}[ \sin(k_1^>x +\frac{2\pi y}{\lambda})- \sin(k_1^<x -\frac{2\pi y}{\lambda})] + 
   \sin[(k_1+k_1^<)x -\frac{2\pi y}{\lambda}]-\sin[(k_1+k_1^>)x +\frac{2\pi y}{\lambda}]\}.
   \label{evanescent}
   \end{eqnarray}
   \endwidetext
  In order to obtain an analytical expression for the spin density, we expand Eqs.\ (\ref{propagating}) and (\ref{evanescent}) with respect to 
  $2\pi x/\lambda$, the actual small parameter being $x/ \sqrt{\lambda \lambda_F}$:
      \widetext
       \begin{eqnarray}
     <S_z(x,y)>=  -\zeta(y)\cdot \frac{\partial S_{str}(x)}{\partial x} + \frac{eV}{(2\pi)^2mv_F^2}(\frac{2\pi x}{\lambda})\cdot W \cos(\frac{2\pi y}{\lambda})\Bigg \{ \int_0^{k_+}dk_y k_y^2 [ \frac{(k_1-k)^2}{kk_1}\cos(k-k_1)x \nonumber \\
      -\frac{(k+k_1)^2}{kk_1}\cos(k+k_1)x  
     +2\cos2kx +2\cos2k_1x ]+  \int_{k_+}^{k_-}dk_yk_y^2 [2\cos2k_1x -2e^{-\kappa x}(\cos k_1x+\frac{k_1}{\kappa}\sin k_1x)] \Bigg \}.
    \label{answer}
   \end{eqnarray}
   \endwidetext
     We note that $S_{str}(x)$ describes the spin density distribution for the straight boundary, see Ref. \cite{Zyuzin}. Therefore, the first term in Eq.\ (\ref{answer}) reflects a small variation of the boundary shape.  The corrected solution $S_{str}(x-\zeta(y))$ contains only $2k_F$ -oscillations.  Let us discuss now the next terms in Eq.\ref{answer} proportional to $1/\lambda$.  As it is clear from Eq.\ \ref{answer}, in general there are oscillations with three different periods, namely,  $2k_F$- oscillations,  and  the oscillations with two long periods: $\xi$ and $L_{s}$. Here $\xi=1/\sqrt{k_-^2-k_+^2}$ is the new length scale, and under the conditions considered in the paper we have the set of inequalities $\lambda_F \ll \xi \ll L_{s}$, where $L_{s}= 1/(k_- - k_+)=\hbar/m\alpha$ is the spin precession length. 
 When $k_y \rightarrow k_+$ (i.e. $k \rightarrow 0$), then $k_1$ tends to $1/\xi$, which clarifies the physical meaning of $\xi$. 
 
    We calculate now the integrals entering expression Eq.\ \ref{answer}
  $$
  I_{1,2}=\frac{1}{k_F}\int_0^{k_+}dk_y k_y^2 \frac{(k_1 \mp k)^2}{kk_1}\cos(k \mp k_1)x,  
  $$
 where the upper (minus) sign corresponds to $I_1$ and the lower (plus) sign to $I_2$. Using new variables (see \cite{Integrals}), we obtain
   \begin{eqnarray}
  I_1= \int_{1/L_{s}}^{1/\xi}dz\cdot \sqrt{z^2-(1/L_{s})^2}\cos(xz)= \nonumber  \\
 = \frac{\sin(x/\xi)}{x\xi}+\frac{\cos(x/\xi)}{x^2}+
  \frac{\pi}{2xL_{s}}N_1(x/L_{s}),
  \label{I_1}
  \end{eqnarray}
  where $N_1(x)$ is the Bessel function of the second kind, and   
  $$
  I_2=4k_F^2\int_{1/2k_F\xi}^{1}dz z \sqrt{1-z^2}\cos(2k_Fxz). 
  $$
Since $1/2k_F\xi \ll 1$,  we can separate the contributions of the long and short wave length oscillations for the propagating modes in Eq.\ \ref{answer}
\widetext
   \begin{eqnarray}
    <S_z(x,y)>= \frac{eV}{(2\pi)^2\hbar v_F}(\frac{2\pi x}{\lambda})\cdot W\cos(\frac{2\pi y}{\lambda})
    [I_1 + I_{1/\xi}+ I_{2k_F}], \nonumber \\
    I_{1/\xi}=4k_F^2\int^{1/2k_F\xi}_{0}dz \cdot z \cos(2k_Fxz)= 
    \frac{\sin(x/\xi)}{x\xi}-\frac{1-\cos(x/\xi)}{x^2},  \nonumber \\
    I_{2k_F}= 2\int_0^1dz \cdot z \sqrt{1-z^2}\Bigg [\frac{k_+^3}{k_F}\cos(2k_+xz)+ \frac{k_-^3}{k_F}\cos(2k_-xz)
   - 2k_F^2\cos(2k_Fxz)\Bigg ]
\label{answer1}
\end{eqnarray}
\endwidetext
Indeed, one  can easily see that the integral $I_{2k_F}$ contains only short wave length oscillations with $2k_F$  momentum, which may be disregarded. 
Therefore,  the total contribution of the {\it long} wave length oscillations is given by
\begin{eqnarray}
    <S_z(x,y)>= \frac{eV}{(2\pi)^2\hbar v_F}(\frac{2\pi W}{\lambda})\cos(\frac{2\pi y}{\lambda})I_{tot}(x), \nonumber \\
   I_{tot}(x)= \frac{2\sin(\frac{x}{\xi})}{\xi}+\frac{2\cos(\frac{x}{\xi})}{x}+
  \frac{\pi}{2L_{s}}N_1(\frac{x}{L_s})-\frac{1}{x} \nonumber \\
+\frac{2x}{\xi}\frac{\partial}{\partial x} \int_0^{1}dze^{-(x/\xi) z} \cos\frac{x\sqrt{1-z^2}}{\xi}, 
\label{total}
\end{eqnarray}
where the last term is the contribution of the evanescent modes. 
At the distances $x\ll \xi$ we obtain the following dependence 
$I_{tot}=-2x^2/3\xi^3 + (x/2L_{s}^2)(\gamma +\ln(x/2L_{s}))$. 
In the opposite limit, $x\gg \xi$, we  obtain
$I_{tot}=\frac{\pi}{2L_{s}}N_1(x/L_{s})-\frac{1}{x}+ \frac{2\cos(x/\xi)}{x}$.  
At even larger distances, $x \gg L_{s}$, one obtains  smooth oscillations with the period of the order of $L_{s}$ and the amplitude being proportional to $\sqrt{\alpha}$. 
\par
 It should be noted that the term $-1/x$ in Eq.\ (\ref{total}) has its origin in the interference of the incoming and outgoing waves, within each mode, 
with $k \rightarrow 0$ and $k_1 \rightarrow 0$. Those waves travel almost parallel to the boundary, and have the scattering angle close to zero. 
Note that the total spin per unit length along the boundary is proportional to the integral $\int dx I_{tot}(x)\simeq (\pi/2L_s)\int_{\xi}^{\infty}dx N_1(x/L_s)-\int_{\xi}^{\sqrt{\lambda \lambda_F}}dx/x \simeq -\ln(L_s/\xi)-\ln(\sqrt{\lambda \lambda_F}/\xi)\simeq -(1/2)\ln(\lambda/\lambda_F)$, i.e., it
is not small in s-o coupling (which can be also seen directly from Eqs.\ (\ref{propagating}) and (\ref{evanescent})), in contrast to the case of the straight boundary.  

\par 

In conclusion, we have considered the problem of the edge spin accumulation in the ballistic mesoscopic structures with spin-orbit-related splitting of the energy spectrum. In the presence of the charge current the spin polarization oscillates at the edge of the structure, as a result of the spin precession upon scattering off the boundary. The scattering states method, being more transparent compared to the methods used earlier by various groups, has allowed us to clarify the physics associated with this effect. We have found that the boundary conditions play a crucial role, even the shape of the sample boundary being important.  In particular, in the case of a straight boundary the absence of spin density oscillations with the long-wave length, found in earlier theoretical works, can be easily explained by the unitarity of scattering. On the other hand, the spin density accumulation in the case of a wiggly boundary has qualitatively different character as compared to the one for the case of a straight boundary, namely, 
the spin density oscillates with a large period of the order of the spin precession length, quite similar to the case  of electron scattering  off antidots in 2D system with the Rashba Hamiltonian \cite{Khaetskii1}. 
Our analysis suggests that a smooth disorder at the boundary should induce spin oscillations with the long wave length. It also implies that in the ballistic case there is no relation between the spin current in the bulk and the spin accumulation at the edge of an electron system.  

 \par 
 We acknowledge the financial support from the Program "Spintronics" of RAS, the Russian Foundation for Basic Research (Grant No. 07-02-00164-a) and the Swiss National Science Foundation.
 A. Khaetskii acknowledges also a hospitality of the Department of Theoretical Physics, University of Geneva.


\begin{thebibliography}{99}
\bibitem{2Dcase}
 As has been mentioned above, only in the effective 1D situation the conditions of the unitarity of scattering and the absence of long-wave oscillations of the spin density coincide. This is no longer the case for 2D systems. Indeed, for a 2D system with the Rashba Hamiltonian the scattering problem of electrons off a circular scatterer (an antidot) may be solved exactly in the presence of the charge current. Then, the condition of the unitarity of scattering has completely different form as compared to the one for the case of scattering 
off the straight boundary. Therefore, $S_z$-component of the spin density, oscillating with period $L_s$, appears around the antidot and decays as a power law of the distance from 
it \cite{Khaetskii1}.
\bibitem{evanes} When $k_+<k_y<k_-$,  then 
 for the reflected plus mode the wave vector in x-direction is purely imaginary. This mode is called evanescent.  
Of course, this situation only happens for the incident minus mode, since only this mode can come from the leads 
under the conditions indicated.
\bibitem{Integrals}
To calculate $I_1$ we introduce the new variable $z=k_1-k=\sqrt{k^2+(1/\xi^2)}-k$, which varies in the interval $1/L_{s}\leq z \leq 1/\xi$. In terms of this variable the quantity $k_y=\sqrt{k_+^2-k^2}=\sqrt{z^2L_{s}^2-1+2\xi^2z^2-\xi^4z^4}/2z\xi^2$.
  Since $\xi \ll L_{s}$, we always have $\sqrt{k_+^2-k^2}= k_+\sqrt{z^2-(1/L_{s})^2}/z$. Note, that $\xi^2k_F=L_{s}/2$. Finally, using that $k_ydk_y=-kdk$, and $dz=-z\cdot dk/k_1$, we obtain the equation of the main text.  To calculate $I_2$  we introduce the variable  $z_1=k_1+k=\sqrt{k^2+1/\xi^2}+k$, which varies in the interval $1/\xi \leq z_1 \leq (k_+ +k_-)=2k_F$. Since $\xi \ll L_{s}$, one gets for the quantity $k_y=\sqrt{k_+^2-k^2}= (1/2)\sqrt{4k_F^2-z_1^2}$. Using that $dz_1=z_1\cdot dk/k_1$, and introducing the variable $z=z_1/2k_F$, we obtain
the expression given in the text.
\end{thebibliography}
\end{document}